\documentclass[pdflatex,sn-mathphys]{sn-jnl}
\usepackage{xcolor}

\jyear{2021}%

\theoremstyle{thmstyleone}%
\theoremstyle{thmstyletwo}%

\theoremstyle{thmstylethree}%

\begin{document}

\title[Article Title]{Atom interferometers and a small-scale test of general relativity}


\author*[1,2]{\fnm{Mikolaj} \sur{Myszkowski}}\email{mikolaj.myszkowski@hertford.ox.ac.uk}

\affil*[1]{\orgdiv{Department of Physics}, \orgname{University of Oxford}, \orgaddress{\street{Parks Road}, \city{Oxford}, \postcode{OX1 3PJ}, \country{United Kingdom}}}


\abstract{Since the first appearance of general relativity in 1916, various experiments have been conducted to test the theory. Due to the weakness of the interactions involved, all of the documented tests were carried out in a gravitational field generated by objects of an astronomical scale. We propose an idea for an experiment that could detect purely general-relativistic effects in a lab-generated gravitational field. It is shown that a set of dense rapidly-revolving cylinders produce a frame-dragging effect substantial enough to be two orders of magnitude away from the observable range of the next generation of atomic interferometers. The metric tensor due to a uniform rotating axisymmetric body in the weak-field limit is calculated and the phase shift formula for the interferometer is derived. This article is meant to demonstrate feasibility of the concept and stimulate further research into the field of low-scale experiments in general relativity. It is by no means a fully developed experiment proposal.}




\maketitle

\section{Introduction}\label{sec1}

The theory of general relativity was first proposed by Einstein in \cite{Article_1}, together with the so called "three classical tests" of the theory: precession of the perihelion of Mercury, the deflection of light by the Sun and gravitational redshift. The irregularity with Mercury's orbit was recognised long before the first appearance of general relativity \cite{Article_41}, while other tests followed after Einstein's article \cite{Article_3,Article_4,Article_2}. Since then, other ways to test general relativity have been found \cite{Article_14, Article_15, Article_16, Article_20}, many of which use solely earth-based apparatus \cite{Article_5, Article_17, Article_18, Article_19}. Significant efforts have been made toward developing interferometers capable of detecting gravitational waves in various frequency bands \cite{Article_60, Article_59, Article_61}. One particularly interesting idea is using atomic interferometry instead of a light-based approach, despite the latter being a widely-used tried and tested method. Examples include the AION, MAGIS, ZAIGA and ELGAR collaborations, whose primary aim is the construction of sufficiently sensitive interferometers, capable of detecting the subtle effects of gravitational waves. \cite{Article_49, Article_50, Article_51, Article_57}.\\\indent
Alternative approach to testing general relativity is via measuring the so called frame-dragging effect. It is predicted that inertial frames should experience "dragging" close to a rotating body, a phenomenon that is absent from Newtonian gravity \cite{Article_85}. Similarly to gravitational waves, the effect is relatively weak and requires very precise apparatus to be observed. Numerous measurements of the effect have been made up to date, both in \cite{Article_81,Article_83,Article_84} and outside of the solar system \cite{Article_82}. In particular, gravity probe B launched in 2004 was able to measure the frame-dragging drift rate of ${-37.2\pm 7.2 \mathrm{mas/yr}}$, in good agreement with the theoretical prediction of ${-39.2\mathrm{mas/yr}}$ \cite{Article_13}.\\\indent
Even though various experiments were conducted, most traditional literature focuses on  measuring the effects of a gravitational field due to large scale objects (e.g. planets or binary systems) rather than human-scale generators. The most promising route toward simultaneous generation and detection of the relativistic effects of gravitation seems to be generation of high-frequency gravitational waves using a strong electromagnetic field \cite{Article_6, Article_7, Article_8, Article_9, Article_11, Article_12}. When high power laser pulse is passed through a dielectric crystal, the energy density inside the medium becomes time dependent. Since the gravitational field couples directly to the stress-energy tensor, the varying energy density distorts nearby gravitational field and creates a high-frequency gravitational wave.\\\indent 
The major drawback of the method described above is the low amplitude ${h\approx 10^{-31}}$ of the resulting gravitational wave \cite{Article_78}. This can be explained by the fact that energy of electromagnetic waves is significantly lower than that resulting from the mass-energy equivalence for massive bodies. The relatively low amplitude still poses a challenge for modern gravitational wave detectors \cite{Article_10, Article_58} that have the goal sensitivity of order ${h\approx 10^{-24}}$ \cite{Article_79}. Despite advances in the field \cite{Article_76}, there is little hope that the gap of order ${10^{-7}}$ between generation and detection capabilities will be resolved in near future.\\\indent
On the other hand, considerable improvements have been made in the field of low-scale tests of Newtonian gravity \cite{Article_43}. Since the famous Cavendish experiment \cite{Article_63}, the universal gravitational constant and the gravitational inverse square law were tested in both short-distance and low-mass regime \cite{Article_42, Article_22}. Only recently, a group of scientists at the Institute for Quantum Optics and Quantum Information in Vienna were able to measure the strength of the gravitational interaction between two gold spheres whose masses were under 100mg \cite{Article_21}.\\\indent
Therefore a question arises as to whether there exist other alternatives that would allow us to observe purely general relativistic effects due to a lab-generated gravitational field. One of the advantages of carrying out measurements with lab-based generators is the strict control over the entire experimental environment. All of the parameters can be adjusted if needed, hence the theory can be examined in a wide range of configurations. Furthermore, general relativity could be tested in the, so far unexplored, short-distance regime.\\\indent 
In this article, we propose an experimental setup that may be capable of observing the frame-dragging due to revolving cylinders of a relatively small scale. The cylinders are placed in the vicinity of an atom interferometer, which fulfills the role of a gravitational field detector. It turns out that the next generation atom interferometers can be a feasible way of probing general relativity in the small-scale regime, provided the noise introduced by the cylinders does not significantly impact their detection capabilities. Atoms traveling in the interferometer experience additional velocity-dependent force due to the external gravitational field generated by the cylinders. Since the effect depends on the velocity of atoms, two atom ensembles moving at different speeds near the cylinders will be affected differently.
The difference between the trajectories of the two beams will result in a phase difference ${\delta \phi}$ in the wave functions of the atoms, which in turn can be accurately measured by the detector.\\\indent
We focus on the relativistically induced contribution to phase shift, arising as a direct result of the frame-dragging generated by the cylinders. The calculations include the phase shift originating from not only the presence of massive cylinders next to the atomic interferometer, but also their non-zero angular velocity. A derivation of the full phase shift formula would require a very thorough analysis of all the details of the experiment, which is outside the scope of the paper. The breakdown of all relevant non-gravitational factors has been already done before, and can be found in \cite{Article_64, Article_65}. Instead, it is shown that by utilizing the precision of next-generation apparatus, the goal of measuring the phase shift due to the frame-dragging effect of the cylinders could be attainable in the near future.\\\indent
We derive the stress-energy tensor of a rotating axisymmetric mass, assuming small angular velocities and no plastic deformation due to centrifugal forces. Furthermore, the weak-field limit of the metric tensor around the rotating axisymmetric body is calculated. We show that in the case of a uniform cylinder, the solution can be partially simplified in terms of integrals of elliptic functions. By considering the effects of the cylinders, the expression for the phase shift in the interferometer is derived. By optimizing the parameters of the generator, next-generation apparatus may be able to capture the phase shift sufficiently accurately for our purposes. Although much emphasis is put on the instructive case of a cylindrically-shaped mass, the results easily generalize to the case of any axisymmetric rotating body.\\\indent
We start by describing the experimental setup, as well as discussing possible sources of noise and ways to reduce it. This is followed by theoretical considerations and a quantitative description of the gravitational field around the generator. Finally, the feasibility of the experiment is assessed and possible improvements are discussed. Throughout the article, we adopt the ${(-,+,+,+)}$ metric signature and ${{\epsilon}_{123}=1}$ sign convention for the Levi-Civita symbol.
\section{Experimental setup}
The underlying idea behind atom interferometry is the same as behind standard light interferometers: particles are split into two groups following different paths. The two beams are then reunited again at the detector, where the phase shift between their wave functions can be measured. Nevertheless, the use of atoms instead of photons as the underlying particles offers numerous advantages. Unlike photons, atoms are not bound to move at the speed of light, which allows for a longer interrogation time (the undisturbed evolution time) of the beam.\\\indent
\begin{figure}[h]
\center
\includegraphics[width=0.8\textwidth]{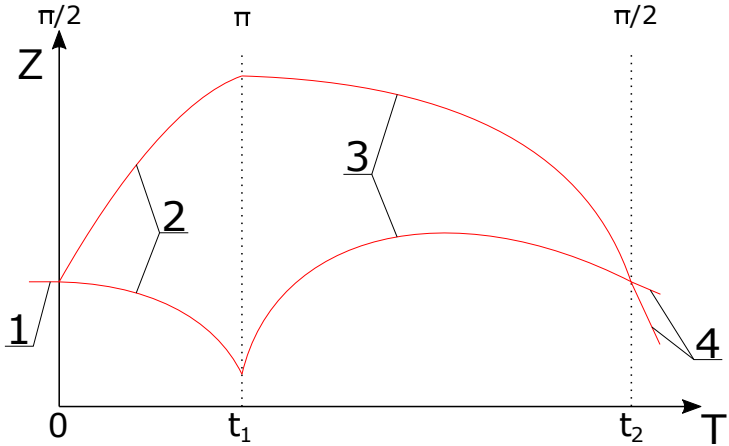}
\caption{\label{fig:fig_1} The spacetime diagram of the classical atom cloud in the Mach-Zehnder atom interferometer. The trajectories of the atoms are represented by a red line, while the light bursts correspond to dotted lines. The ensemble of atoms enters the chamber with a vanishing initial velocity (1). Half of the atoms pick up a vertical momentum ${\hslash k_{eff}}$ from the photon beam via the ${\pi /2}$ Bragg transmission (2). After the time ${t_{1}}$ elapses, they experience a ${\pi}$ Raman pulse, gaining momenta of ${-\hslash k_{eff}}$ and ${\hslash k_{eff}}$ for the upper and lower group of atoms respectively (3). Finally the atoms are reunited with a second Bragg pulse and reach the detector (4).}
\end{figure}

The complication is that atoms cannot be redirected using regular mirrors and beam splitters. Instead, entire groups of atoms are manipulated using short bursts of light. By tuning the laser's parameters, a coherent momentum transition between light and atoms is possible \cite{Article_62}, thus allowing for consistent manipulation of whole atom ensembles. The spacetime diagram for the atoms traveling vertically in the interferometer's arm is shown in Fig.~\ref{fig:fig_1}. The time evolution of the wave function can be divided into four separate stages.\\\indent
Firstly, the cloud of ultracold atoms created in an atom source is introduced into the vacuum chamber of the interferometer (1). The first stage is then followed by a short pulse of light at ${T=0}$, during which some of the atoms are Bragg scattered and gain vertical momentum ${\hslash k_{eff}}$, where ${k_{eff}=2k}$ is double the wavenumber of the laser beam. The factor of two originates from the fact that the atom firstly absorbs a photon with momentum ${p=\hslash k}$ and then emits a photon with momentum ${p=-\hslash k}$ \cite{Article_66}.\\\indent
The wave function of the ensemble is now a superposition of two states: the one that interacted with the photons (the upper red line) and the non-disturbed state, which is in a free-fall due to earth's gravitational field (2). At the time ${T=t_{1}}$, both states go through Raman transitions, with the upper state losing momentum, and the lower state gaining momentum in the process. The overall change in the vertical momentum for each atom ensemble is once more double the photons' momentum ${\hslash k_{eff}}$ (3). Finally, the two states overlap again at the time ${T=t_{2}}$, after which they are reconciled with another Bragg pulse (4) and redirected towards the detector.\\\indent
The final state is a superposition of the upper and lower wave functions. Since the ensembles were evolving over two different paths, any gravitational effects acting unevenly along the trajectories lead to a potential phase shift between the upper and lower wave functions, which can be measured to a high accuracy at the detector.\\\indent
\begin{figure}[h]
\center
\includegraphics[width=0.4\textwidth]{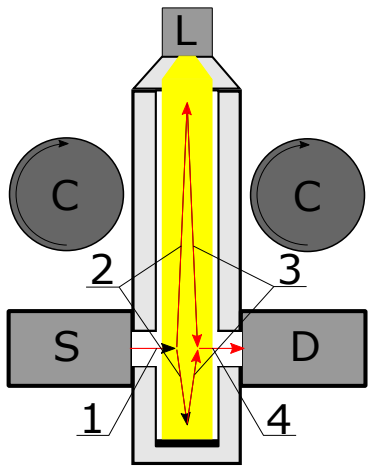}
\caption{\label{fig:fig_2} Schematic representation of the experiment. The ensemble of ultracold atoms (red arrow) is prepared in the atom source (S) before entering the main vacuum chamber. The laser then interacts with the atomic cloud via a combination of Raman and Bragg transmissions, as shown in Fig.~\ref{fig:fig_1}. The two counter-propagating pulses of light needed for manipulation of the atoms are generated using the clock laser apparatus (L) together with the reflector at the bottom of the vacuum chamber (bold black line). Finally, the two atom clouds reunite again and are redirected horizontally towards the detector. The phase shift between the wave functions of the two ensembles can be observed in the detector (D) using the imaging of clouds method. The entire process takes place in the external gravitational field generated by the earth and rotating cylinders (C). The cylinders are arranged symmetrically with respect to the centre of the vacuum chamber. For simplicity, the subsystems required to run the apparatus and control the cylinders' spin are not indicated in the picture.}
\end{figure}
The full experimental setup including the cylinders is illustrated in Fig.~\ref{fig:fig_2}. The cylinders are responsible for the creation of the frame-dragging effect, while the atom interferometer plays the role of the detector. Firstly, a cloud of atoms (typically ${{}^{85}\mathbf{Rb}}$ or ${{}^{87}\mathbf{Rb}}$) is prepared in the atom source (S) by laser cooling to low temperatures of order ${10^{-5}\mathrm{K}}$ \cite{Book_8}. At this stage, the movement of the ensemble is restricted by a magneto-optical trap. Usually, the final flux of atoms in the ensemble is kept low enough (${\sim 10^{6}\mathrm{s^{-1}}}$) so that any interactions between atoms can be neglected. The atoms are then released into the main vacuum chamber and manipulated with use of the clock laser (L). The four sections of the trajectory (1), (2), (3) and (4) correspond to the same four stages indicated in Fig.~\ref{fig:fig_1}. Following the series of laser pulses, the cloud is redirected back to the detector, where the potential phase shift can be measured \cite{Article_68}. Because of the nature of the setup, atoms need to be redirected horizontally towards the detector at the end of the interferometer sequence. This can be achieved with an additional laser pulse that, unlike the pulses in Fig.~\ref{fig:fig_1}, would be directed non-vertically.\\\indent
The accuracy of the phase shift obtained increases with the number of atoms in the ensemble and the integration time (the time over which the measurements are taken). Modern interferometers operate in the phase sensitivity range of ${10^{-3}\mathrm{rad}}$ to ${10^{-4}\mathrm{rad}}$, which is more than enough to measure very subtle phase changes due to the gravitational field \cite{Article_48}. The next generation of interferometers currently under development (e.g. AION100) intend to push the limits further, being capable of measurements accurate to ${3\cdot 10^{-6}\mathrm{rad}}$ \cite{Article_49}.\\\indent
The force corresponding to the frame-dragging effect due to rotation of the cylinders depends on two factors: the velocity of the test particle and its distance from the generator. The magnitude of the cylinder-induced gravitational effects diminishes as the particle moves slower and further away from the cylinders. Similarly, the force changes direction if the velocity of the particle reverses. Therefore, a non-zero separation is expected between the wave functions of the upper and lower atom ensembles, leading to a measurable phase shift at the detector.\\\indent
In order to attain the required levels of accuracy from the setup, any noise (from the environment, the cylinders or otherwise) should be minimised. Noise from the cylinders is especially important, as the massive fast-rotating cylinders together with corresponding subsystems can be a major source of vibration noise. In order to minimize potential impact on the interferometer, the cylinder-related setup should be properly isolated from the detecting apparatus. Possibly, a magnetic suspension of the cylinders could be employed to reduce any periodic vibrations due to the cylinders' imperfections \cite{Article_69}. The external noise can be significantly lowered by shielding the interferometer from electromagnetic background, as well as installing the detection apparatus on an active vibration isolation system \cite{Article_70}.\\\indent
Besides the random phase difference arising purely from environmental noise, other effects will appear. Earth's rotation will create an additional systematic phase shift due to the Coriolis effect, which is dependent on the location of the experiment on the globe \cite{Article_71}. An additional change in phase can be expected due to the gradient of the gravitational field along the length of the vacuum chamber. One of the advantages of measuring the effects of the cylinders' gravitational field rather than the external field is that the angular speed of the cylinders can be modulated. With increased angular velocity, the phase shift due to the frame-dragging effect will increase, while the contributions from other systematic errors should stay approximately the same. Instead of measuring the absolute value of the phase shift, a series of measurements can be taken and the change of the phase shift with the angular frequency of the cylinders can be examined.\\\indent
Although the above remarks outline key sources of noise, it is by no means a fully developed analysis of the issue. Various other factors have to be considered, including laser-related noise and the shot noise. A thorough review of the topic for the current generation of interferometers can be found in \cite{Article_67, Book_8}.
\section{Theoretical analysis}
The calculation of the phase shift can be split into two sections. Firstly, the stress-energy tensor due to an axisymmetric rotating mass distribution is calculated. We assume that the body does not deform under centrifugal forces, and that the stress-energy tensor can be approximated by that of a perfect fluid. The weak-field metric tensor is then obtained in the case of a uniform cylinder, which can be written in terms of integrals of elliptic functions.\\\indent
Secondly, the dynamics of atom ensembles are considered. Due to the symmetric properties of the system, atoms are bound to move in the plane of cylinders throughout the interferometer sequence (Fig.~\ref{fig:fig_1}). Therefore, calculations simplify and the problem can be reduced to that of a 1+2 dimensional test particle. Because the cylinders rotate, there exist a small velocity-dependent correction to the classical motion of the ensembles. Hence, the trajectories can be well-approximated by the classical
paths (derived from Newton’s equations) perturbed by the weak gravitational field of the cylinders.\\\indent
In general, the total phase shift constitutes three main contributions: The propagation component (the phase difference accumulated throughout the free evolution of the atom ensembles), the laser component (the phase shifts due to the laser-atom interactions) and the separation component (the phase shift due to the resulting separation of the ensembles at the detector). Due to the nature of the velocity-dependent correction to the ensembles' motion, laser and propagation terms can be disregarded. Finally, a compact formula for the phase shift is derived. The result is then applied to estimate the expected magnitude of the phase shift for a modern interferometer of length ${10\mathrm{m}}$.
\subsection{Spacetime around a rotating axisymmetric mass distribution}
\begin{figure}[h]
\center
\includegraphics[width=0.5\textwidth]{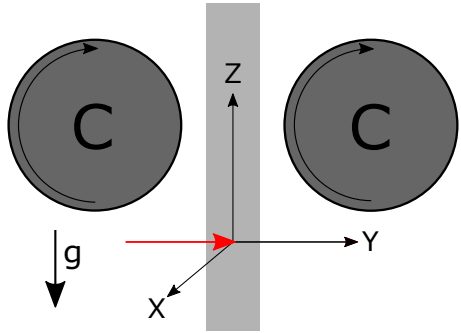}
\caption{\label{fig:fig_3} The set of cylinders (C) embedded in 3D Cartesian coordinates. The arm of the interferometer is represented in light grey for better reference. The ${z}$ axis is parallel to the vacuum chamber and the direction of the gravitational field. Conveniently, the atom injection point (the end of red arrow) overlaps with the origin ${(t,x,y,z)=(0,0,0,0)}$ of the coordinates. The centres of the cylinders lie in the ${y-z}$ plane.}
\end{figure}

There are two contributions to the stress-energy tensor of a rotating body: the energy of the cylinder itself and of the force responsible for keeping the cylinder together. However, in the lower speed limit, the latter is expected to be negligible, leaving only the dominating kinetic term. Therefore, the stress-energy tensor takes the familiar form of a perfect fluid:
\begin{equation}
T_{\mu \nu}=\rho u_{\mu}u_{\nu}
\end{equation}
where ${\rho =\rho(r,z)}$ is the mass density of the body. In cylindrical coordinates ${(t,r,\theta,z)}$ the flat metric tensor reads:
\begin{equation}
\eta_{\mu \nu}=diag\left[-1,1,r^{2},1\right].
\end{equation}
For simplicity, the symmetry axis of the cylinder is assumed to overlap with the ${z}$ axis. The four-velocity of a body rotating at angular velocity ${\omega}$ is given by:
\begin{equation}
u^{\mu}=\gamma (c,0,\omega,0), \ \gamma =\left(1-\frac{{\omega}^{2}r^{2}}{c^2}\right)^{-1/2}
\end{equation}
leading to a kinetic contribution of the form:
\begin{equation}
T_{\mu \nu}^{cyl}=\rho u_{\mu}u_{\nu}={\gamma}^2\rho c^{2}
\begin{bmatrix}
1 & 0 & -r^{2}\frac{\omega}{c} & 0 \\
0 & 0 & 0 & 0 \\
-r^{2}\frac{\omega}{c} & 0 & r^{4}\frac{{\omega}^{2}}{c^2} & 0 \\
0 & 0 & 0 & 0 \\
\end{bmatrix}
\end{equation}\\\indent
In practical applications, the velocity of the rotating body is small compared to the speed of light, i.e. ${r\omega<<c}$. The leading term in the stress-energy tensor is the energy term ${T_{00}\approx \rho c^{2}}$, as expected.
However, the phenomenon of frame-dragging is induced by the appearance of the non-diagonal terms \cite{Book_1}, hence the ${T_{02}, T_{20}}$ components have to be kept as well. By taking into account contributions of order ${\sim r^{2}\omega/c}$, we arrive at the low angular velocity limit of the stress-energy tensor in Cartesian coordinates ${(t,x,y,z)}$:
\begin{equation}
T_{\mu \nu}^{cyl}\approx \rho c^{2}
\begin{bmatrix}
1 & 0 & -r^{2}\frac{\omega}{c} & 0 \\
0 & 0 & 0 & 0 \\
-r^{2}\frac{\omega}{c} & 0 & 0 & 0 \\
0 & 0 & 0 & 0 \\
\end{bmatrix} \implies \ 
T_{\mu \nu}^{crt}\approx \rho c^{2}
\begin{bmatrix}
1 & \frac{\omega}{c}y & -\frac{\omega}{c}x & 0 \\
\frac{\omega}{c}y & 0 & 0 & 0 \\
-\frac{\omega}{c}x & 0 & 0 & 0 \\
0 & 0 & 0 & 0 \\
\end{bmatrix}
\end{equation}
The weak-field approximation of the metric ${g_{\mu \nu}^{crt}={\eta}_{\mu \nu}+h_{\mu \nu}}$ can be obtained by integrating over all infinitesimal contributions from the cylinder:
\begin{equation}
{\tilde{h}}_{\mu \nu}\left(\vec{\bf x}\right)=\frac{4G}{c^{4}}\int_V\frac{T_{\mu \nu}^{crt}\left(\vec{\bf x}'\right)}{\lvert \vec{\bf x}-\vec{\bf x}'\rvert}dV'
\end{equation}
where ${h_{\mu \nu}={\tilde{h}}_{\mu \nu}-\frac{1}{2}\tilde{h}{\eta}_{\mu \nu}}$ \cite{Book_1}.\\\indent
The ${{\tilde{h}}_{00}}$ term is proportional to the Newtonian potential, and thus for a stationary cylinder can be expressed in terms of elliptic integrals. The Newtonian potential ${\Phi}$ outside an infinitesimal disc of radius ${R}$ and mass ${M}$ (centered at the coordinate's origin and perpendicular to the ${z}$ axis) is \cite{Article_25, Article_26}:
\begin{equation}
\begin{aligned}
\Phi=-\frac{2MG}{\pi R^{2}\sqrt{z^{2}+(R+r)^{2}}}\bigg [\left(z^{2}+(R+r)^{2}\right)E(k) \\
+\left(R^{2}-r^{2}\right)K(k)+\frac{(R-r)z^{2}}{(R+r)}\Pi (n,k)\bigg],
\end{aligned}
\end{equation}
where:
\begin{equation}
\begin{aligned}
&r=\sqrt{x^2+y^2}, \\
& k^{2}=\frac{4Rr}{z^{2}+(R+r)^{2}}, \ n=\frac{4Rr}{(R+r)^{2}}, \\
& K(k)=\int^{1}_{0}\frac{dt}{\sqrt{(1-t^{2})(1-k^{2}t^{2})}}, \\
& E(k)=\int^{1}_{0}\frac{\sqrt{1-k^{2}t}}{\sqrt{1-t^2}}dt, \ \ and \\
& \Pi (n,k) =\int^{1}_{0}\frac{dt}{(1-nt^{2})\sqrt{(1-t^{2})(1-k^{2}t^{2})}}.
\end{aligned}
\end{equation}
Therefore, by integrating over infinitesimal discs, the ${\tilde{h}_{00}}$ term of the linearized metric is obtained:
\begin{equation}
\tilde{h}_{00}(r,z)=\frac{4}{c^{2}H}\int_{-z-H/2}^{-z+H/2}\Phi (r,z')dz',
\end{equation}
where ${H}$ is the height of the cylinder (the additional factor of ${H}$ in the denominator appears as the mass of infinitesimal disc is given by ${Mdz/H}$, where ${M}$ becomes the total mass of the cylinder) and ${r}$ is given by ${r=\sqrt{x^2+y^2}}$. Since the cylinder is centred at ${z=0}$ and ${\tilde{h}_{00}}$ is evaluated at ${z}$, the potential ${\Phi (r,z')}$ has to be integrated from ${-z-H/2}$ to ${-z+H/2}$.\\\indent
The off-diagonal terms are difficult to calculate with the same method. The difference is that the density of the cylinder is weighted with ${\sim y}$ (or ${\sim -x}$) term, making it hard to express the integral using elliptic functions. In other words, ${{\tilde{h}}_{01}}$ and ${{\tilde{h}}_{20}}$ are proportional to the Newtonian potential due to a cylinder with density ${\rho'=\omega y/c}$ (or ${\rho'=-\omega x/c}$ respectively). The equation (6) can be now expanded:
\begin{equation}
\tilde{h}_{01}\left(\vec{\bf x}\right)=\tilde{h}_{10}\left(\vec{\bf x}\right)=\frac{4G\rho\omega}{c^{3}}\int_{-H/2-z}^{H/2-z}\int_{r\leq R}\frac{y'}{\lvert \vec{\bf x}-\vec{\bf x}'\rvert}dV'
\end{equation}
where ${\vec{\bf x}=(x,y,z)}$ and ${\vec{\bf x}'=(x',y',z')}$. Similarly, the ${\tilde{h}_{02}}$ component may be written as:
\begin{equation}
\tilde{h}_{02}\left(\vec{\bf x}\right)=\tilde{h}_{20}\left(\vec{\bf x}\right)=-\frac{4G\rho\omega}{c^{3}}\int_{-H/2-z}^{H/2-z}\int_{r\leq R}\frac{x'}{\lvert \vec{\bf x}-\vec{\bf x}'\rvert}dV'
\end{equation}
which leads to the final form of the metric in Cartesian coordinates:
\begin{equation}
\begin{aligned}
&g_{\mu \nu}^{crt}={\eta}_{\mu \nu}+\tilde{h}_{\mu \nu}-\frac{1}{2}\tilde{h}{\eta}_{\mu \nu}=\\&
\begin{bmatrix}
-1+\frac{1}{2}\tilde{h}_{00} & \tilde{h}_{01} & \tilde{h}_{02} & 0 \\
\tilde{h}_{10} & 1+\frac{1}{2}\tilde{h}_{00} & 0 & 0 \\
\tilde{h}_{20} & 0 & 1+\frac{1}{2}\tilde{h}_{00} & 0 \\
0 & 0 & 0 & 1+\frac{1}{2}\tilde{h}_{00} \\
\end{bmatrix}
\end{aligned}
\end{equation}
with ${\tilde{h}_{00}}$, ${\tilde{h}_{01}=\tilde{h}_{10}}$, and ${\tilde{h}_{02}=\tilde{h}_{20}}$ defined as in (9), (10) and (11) respectively.\\\indent
The metric tensor (12) is expressed in coordinates where the cylinder is placed at the origin. In order to calculate the metric around the ensemble of cylinders arranged symmetrically around the interferometer's arm, it is convenient to transform to new shifted Cartesian coordinates. The system of cylinders is depicted in Fig.~\ref{fig:fig_3}. In this new coordinate system, the atoms are injected at the point ${(t,z,y,z)=(0,0,0,0)}$ and throughout the time evolution remain in the ${y-z}$ plane, i.e. ${x=0}$.\\\indent

Let the position of the centre of each cylinder be denoted by ${(t,0,\pm y_{c},z_{c})}$. The perturbation tensor ${h_{\mu \nu}^{\pm}}$ in the newly defined coordinates can now be obtained by transforming (12) to the coordinate system  shown in Fig.~\ref{fig:fig_3}. Tensors ${h_{\mu \nu}^{+}}$ and ${h_{\mu \nu}^{-}}$ denote the contributions from the cylinder at ${(t,0, y_{c},z_{c})}$ and ${(t,0,-y_{c},z_{c})}$ respectively:
\begin{equation}
\begin{aligned}
&h_{\mu \nu}^{\pm}=
\begin{bmatrix}
\frac{1}{2}\tilde{h}_{00} & 0 & \tilde{h}_{01} & -\tilde{h}_{02} \\
0 & \frac{1}{2}\tilde{h}_{00} & 0 & 0 \\
\tilde{h}_{10} & 0 & \frac{1}{2}\tilde{h}_{00} & 0 \\
-\tilde{h}_{20} & 0 & 0 & \frac{1}{2}\tilde{h}_{00} \\
\end{bmatrix}
\end{aligned}
\end{equation}
where the components ${\tilde{h}_{00}}$ and ${\tilde{h}_{20}=\tilde{h}_{02}}$ transform like a scalar, and their arguments change from ${(x,y,z)}$ to ${(y\pm y_{c},z_{c}-z,x)}$. Therefore, the full metric tensor reads:
\begin{equation}
g_{\mu \nu}=g_{\mu \nu}^{bg}+h_{\mu \nu}^{+}+h_{\mu \nu}^{-}=g_{\mu \nu}^{bg}+h_{\mu \nu}^{tot}
\end{equation}
where ${g_{\mu \nu}^{bg}}$ is the background metric in which the interferometer is placed.
\subsection{Relativistic corrections to atoms' dynamic}
The evolution of the atom ensemble is piecewise geodesic, i.e. the atoms are subjected only to gravitational field in between the laser pulses. Considering the geodesics in between the laser interactions, a trajectory can be parametrized by the atom's proper time ${\tau}$, leading to the position four-vector:
\begin{equation}
X^{\mu}=\left(x^{0}(\tau),0,x^{2}(\tau),x^{3}(\tau)\right).
\end{equation}
The placement of the cylinders is mirror symmetric in the ${x^{1}}$ component. This, combined with the fact that the ensemble is assumed to have no initial velocity at entry results in ${x^{1}(\tau)=0}$. In other words, atoms move in the ${y-z}$ plane throughout the interferometer sequence.\\\indent
The full geodesic equations associated with metric (14) are too complicated to be solved analytically. Instead, the leading relativistic correction to the classical path is calculated by working in the Einstein-Maxwell formulation of linearized gravity. As opposed to the regular linearized regime which leads to Newton's equation, the Einstein-Maxwell framework accounts for the off-diagonal terms in the metric tensor. The ${h_{0 \mu}, \mu \neq 0}$ components of the metric manifest themselves through a velocity-dependent force, in analogy to the Maxwell equations. The approximate dynamics of the ensembles can be then examined using the framework of classical mechanics.\\\indent
The background metric can be taken to be the Schwarzschild metric. Nevertheless, atom interferometers are sensitive to variations in earth's mass density and thus correcting terms may sometimes be needed. In addition, only the 1st and 2nd gradient of the gravitational field lead to measurable effects. As a result, it is often more convenient to work with Taylor expanded background gravitational potential:
\begin{equation}
{\Phi}^{bg} \approx c_{0}+c_{1}\frac{z}{L}+c_{2}\left(\frac{z}{L}\right)^{2}
\end{equation}
where ${L}$ is the characteristic length of the interferometer. The coefficients ${c_{0}, c_{1}, c_{2}}$ depend on the localization of the apparatus and other more subtle details. It is assumed that the background metric varies slowly with ${z}$, i.e. we expect ${c_{1}>>c_{2}}$. The background metric corresponding to (16) is:
\begin{equation}
g_{\mu \nu}^{bg}=diag\left[-1-2\frac{{\Phi}^{bg}}{c^{2}},1-2\frac{{\Phi}^{bg}}{c^{2}},1-2\frac{{\Phi}^{bg}}{c^{2}},1-2\frac{{\Phi}^{bg}}{c^{2}}\right].
\end{equation}
The full metric is the sum of the above mentioned expanded background metric and the perturbation terms, as stated in (14).\\\indent 
Since the geodesics are effectively bounded to 1+2D, there is no need to refer to the full geodesic equations and the dynamic of the ensemble can be determined using the velocity normalisation condition and the conservation of energy. The trajectory of an ensemble is defined by a position vector (15) parametrized with the ensemble's proper time ${\tau}$. The associated four-velocity reads:
\begin{equation}
U^{\mu}=(c{\partial}_{\tau} t, 0, {\partial}_{\tau} y, {\partial}_{\tau} z)=({\partial}_{\tau} x^{0}, 0, {\partial}_{\tau} x^{2}, {\partial}_{\tau} x^{3})
\end{equation}
and satisfies the normalization condition:
\begin{equation}
\begin{aligned}
-c^{2}=&\bar{g}_{\mu \nu}U^{\mu}U^{\nu}=(g_{00}^{bg}+h_{00}^{tot})({\partial}_{\tau} x^{0})^{2}+(g_{22}^{bg}+h_{22}^{tot})({\partial}_{\tau} x^{2})^{2}+\\
&(g_{33}^{bg}+h_{33}^{tot})({\partial}_{\tau} x^{3})^{2}+2h_{02}^{tot}({\partial}_{\tau} x^{0})({\partial}_{\tau} x^{2})+2h_{03}^{tot}({\partial}_{\tau} x^{0})({\partial}_{\tau} x^{3})
\end{aligned}
\end{equation}
Since the underlying metric is static and does not depend on the ${x^{0}}$ coordinate, the energy of test particles is conserved. The conservation law can be written using the Killing vector ${{\xi}^{\mu}=(c,0,0,0)}$ associated with time invariance:
\begin{equation}
\frac{E}{m}=-g_{\mu \nu}{\xi}^{\mu}U^{\nu}=-c(g_{00}^{bg}+h_{00}^{tot}){\partial}_{\tau} x^{0}-2ch_{02}^{tot}{\partial}_{\tau} x^{2}-2ch_{03}^{tot}{\partial}_{\tau} x^{3}
\end{equation}
The energy to mass ratio for the atoms ${E/m}$ is predetermined by the atom-laser interactions during the interferometer sequence.\\\indent
Although the equations (19) and (20) specify the dynamics of the ensembles, the non-linearity and complicated forms of ${\tilde{h}_{00}, \tilde{h}_{01}}$ make it difficult to find an explicit solution. Instead, the problem can be reformulated in terms of the Einstein-Maxwell equations. The Einstein-Maxwell framework can be viewed as a more inclusive approximation of general relativity, where the velocity-dependent forces are accounted for. The outline of the method is presented in Appendix A. \\\indent
Similarly to electromagnetism, we define auxiliary fields:
\begin{equation}
{E}_{i}=\frac{c^{2}}{2}{\partial}_{i}h_{00}^{tot} \ and \ {B}_{i}=c{\epsilon}_{ijk}{\partial}_{j}h_{0k}^{tot}
\end{equation}
where ${\bf E}$ and ${\bf B}$ are gravitational analogues of electric and magnetic fields. Then, the acceleration of a test particle moving with velocity ${\bf v}$ due to the gravitational field can be compactly written as:
\begin{equation}
{\partial}_{t}^{2}{\bf x}={\bf E}+{\bf v}\times {\bf B}
\end{equation}

\begin{figure*}
\center
\includegraphics[width=\textwidth]{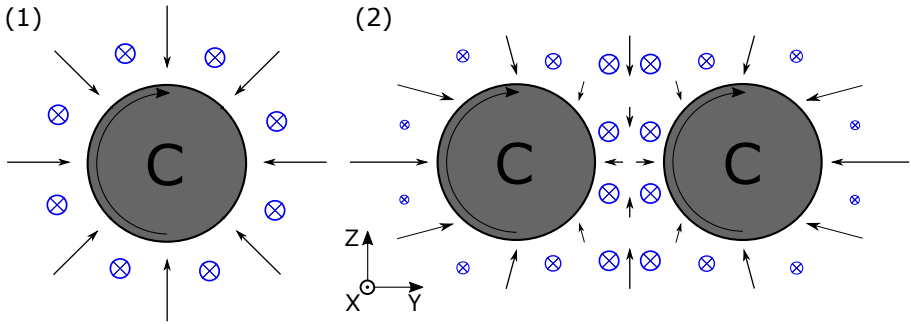}
\caption{\label{fig:fig_4} The gravitational fields around rotating cylinders. The ${\bf E}$ field (black arrows) is directed towards the cylinders as opposed to the ${\bf B}$ field (blue arrows), which is perpendicular to the page. The gravitational fields around a single cylinder are depicted in (1), while the fields around two cylinders rotating in the same direction are shown in (2). In the latter, the total field is just a sum of contributions from each of the cylinders. On the axis of symmetry, ${{{\partial}_{x}\bf B}, {{\partial}_{y}\bf B}}$ vanish and ${\bf E}$ points in the ${z}$ direction only.}
\end{figure*}

The gravitational fields ${\bf E}$ and ${\bf B}$ around the rotating cylinders are shown in Fig.~\ref{fig:fig_4}\\\indent
For reference, a single cylinder rotating clockwise is depicted in (1). The curved black arrow indicates the direction of rotation of the cylinder. ${\bf E}$ plays the role of a regular gravitational field and its direction does not depend on the rotation of the cylinder, as expected. Contrary to ${\bf E}$, ${\bf B}$ is explicitly dependent on the angular speed of the cylinder, i.e. ${{\bf B}\propto \omega}$. The direction of the field is determined by the direction of the rotation. Should the cylinder rotate counter clockwise, the sign of ${\bf B}$ would reverse as well.\\\indent
The image (2) depicts a set of two clockwise rotating cylinders, already shown in Cartesian coordinates in Fig.~\ref{fig:fig_3}. As mentioned earlier, the ${F_{x}}$ component of the gravitational force (22) acting on a test particle that moves in the ${y-z}$ plane vanishes. Due to the symmetry of the system, ${\bf E}$ and ${\bf B}$ fields point in the ${z}$ and ${-x}$ direction respectively:
\begin{equation}
{\bf E}\times {\bf \hat{z}}{\lvert}_{x=y=0}={\bf 0}, \ {\bf B}\times {\bf \hat{x}}{\lvert}_{x=y=0}={\bf 0}
\end{equation}
when evaluated on the axis ${x=y=0}$. Since ${{\bf B}}$ is dependent on the angular velocity ${\omega}$, the trajectories of the ensembles will be affected not only by the gravitational field ${{\bf E}}$, but also the rotation of the cylinders.
\subsection{Derivation of the phase shift formula}
\begin{figure*}
\center
\includegraphics[width=\textwidth]{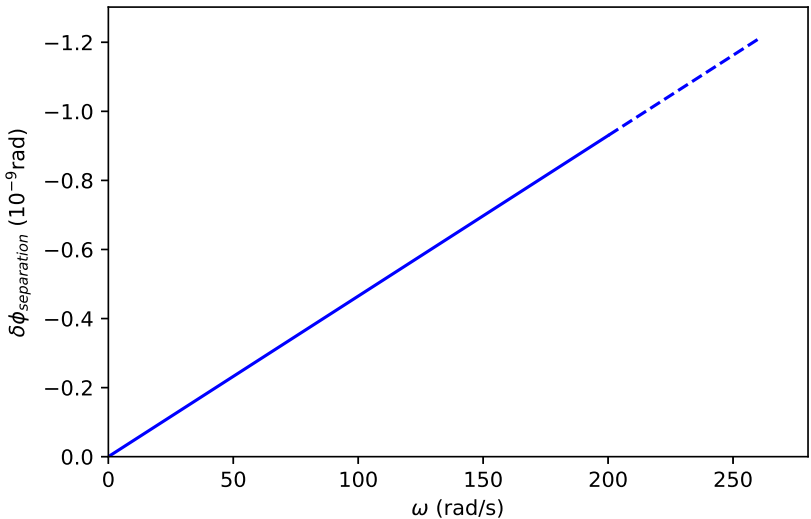}
\caption{\label{fig:fig_5} The separation phase shift ${\delta {\phi}_{separation}}$ plotted as a function of the angular velocity of the cylinders ${\omega}$ for a typical interferometer with ${L\sim 10\mathrm{m}}$, ${\Delta v\sim 14\mathrm{m/s}}$, ${t_{2}=2t_{1}\sim 1.4\mathrm{s}}$ and ${v_{y}\sim 10\mathrm{m/s}}$. For this particular configuration, the cylinders are placed ${4\mathrm{m}}$ above the point of entrance of the ensembles and ${3.15\mathrm{m}}$ away from the interferometer's arm. Each cylinder has width ${0.4\mathrm{m}}$ and radius ${R\sim 3\mathrm{m}}$, which limits the maximal rotational velocity of the cylinders to ${\omega\sim 200\mathrm{rad/s}}$. The separation phase shift increases linearly from zero (when the cylinders do not rotate, and hence no frame-dragging effect is present) to ${\delta {\phi}_{separation}=-9.3\cdot 10^{-10}\mathrm{rad}}$ at the maximal angular velocity ${\omega=200\mathrm{rad/s}}$. The dotted line represents the behaviour of phase shift in the regime ${\omega > 200\mathrm{rad/s}}$, where the centrifugal force exceeds the tensile strength of the cylinders.}
\end{figure*}

The phase difference between the wave functions of the two atom ensembles can be derived in a stationary phase approximation, where atoms are assumed to follow the corresponding point-particle trajectory. This approximation is expected to break down if the gravitational field varies too rapidly within the interferometer's arm. Even though the cylinders are relatively close to the vacuum chamber, the gravitational field of the cylinders is weak enough so that the approximation still holds. An example of this fact is the determination of the gravitational constant ${G}$ using atom fountains. In this type of experiments, a total mass of order ${\sim 10^{4}\mathrm{kg}}$ is placed near the vacuum chamber of the interferometers, and even for the most precise measurements the size of the atomic cloud is negligible \cite{Article_74}.\\\indent
Another important simplification that is used during calculations is the so called  "short pulse limit". We assume the gravitational effects themselves do not change atoms' velocity significantly over the duration of the laser pulse. Therefore, the interaction of light with atoms can be treated as instantaneous (the change of atoms' velocity due to the laser pulse occurs instantaneously), which greatly simplifies the considerations. For very precise measurements, this assumption is known not to hold. However, the corrections can be easily calculated within the Newtonian framework, and do not require a fully general-relativistic treatment. As a result, we choose to omit the derivations of the corresponding well-known effects, which can be found in other works \cite{Article_72, Article_73}. As usual, it is assumed that the flux of atoms is kept low enough so that the inter-atom coupling does not have a significant impact on the result. \\\indent
The trajectories of the atoms are fully parametrized by the times ${t_1}$ and ${t_2}$ of the laser pulses together with the wave number of the photons ${k}$. Nevertheless, it is more convenient to formulate expressions in terms of the classical paths of the ensembles. Analogously to \cite{Article_46}, we define:
\begin{equation}
{\bf x}_{u}(t)=(0,0,z_{u}(t)), \ {\bf x}_{l}(t)=(0,0,z_{l}(t))
\end{equation}
to be the classical trajectories of the lower and upper ensemble respectively. For a particular experiment, the trajectories ${{\bf x}_{l}(t)}$ and ${{\bf x}_{u}(t)}$ are just the paths of the ensembles in the presence of non-rotating cylinders. When cylinders start spinning, the additional velocity-dependent force ${m{\bf v}\times {\bf B}}$ perturbates the upper and lower trajectories by a small amount ${\delta {\bf x}_{u}(t)}$ and ${\delta {\bf x}_{l}(t)}$ respectively. In general, ${{\bf x}_{u}(t)}$, ${{\bf x}_{l}(t)}$, ${\delta {\bf x}_{u}(t)}$ and ${\delta {\bf x}_{l}(t)}$ take rather complicated forms and cannot be given analytically for all but the simplest gravitational potentials. Furthermore, they often depend on the specific experimental setup, and therefore we feel that the most appropriate approach is to derive the phase shift in terms of (24) and their small perturbations ${\delta {\bf x}_{u}(t)}$, ${\delta {\bf x}_{l}(t)}$. When evaluating the phase shift for individual experiments, the trajectories can be obtained
using previously developed techniques (e.g. numerically or using Taylor expansion \cite{Article_46}).\\\indent
The phase shift due to the cylinders' gravitational field can be written as a sum of three different contributions \cite{Article_46}:
\begin{equation}
\delta {\phi}_{tot} =\delta {\phi}_{laser}+\delta {\phi}_{propagation}+\delta {\phi}_{separation}
\end{equation}
Each component in the sum has a different conceptual origin.\\\indent
The laser phase shift is a results of the atom-laser interaction and depends on the laser setup. A detailed derivation of the formula can be found in \cite{Article_46}, and therefore we decide to outline points of particular importance to our considerations. The laser-atom interactions can be modeled by a single two-level atom coupled to a monochromatic laser beam. The electric wave inside the interferometer's arm can be described by:
\begin{equation}
{\bf E}({\bf x},t)={\bf E}_{0}cos({\bf k}\cdot {\bf x}-\omega t+\psi)
\end{equation}
Provided that the short pulse limit holds, the phase shift due to a single atom transmission is just the laser phase at the time of the interaction ${t_{int}}$:
\begin{equation}
\Delta{\phi}({\bf x}_{int},t_{int})=\pm \left({\bf k}\cdot {\bf x}_{int}-\omega t_{int}+\psi\right)
\end{equation}
where ${{\bf x}_{int}}$ denotes the position of the atom at the interaction time ${t_{int}}$. The sign depends on whether the atom loses or gains momentum in the process. If photon is absorbed, the atom gains momentum and the sign is positive. Similarly, if the photon is emitted, atom loses momentum and the sign in front of the phase shift must be negative. The laser phase shift ${\delta {\phi}_{laser}}$ can be then written as a sum over all individual laser-atom interactions during the interferometer sequence (Fig.~\ref{fig:fig_1}). Since the final laser phase shift is the relative phase difference between the upper and lower atom assemblies, ${\delta {\phi}_{laser}}$ is given by:
\begin{equation}
\begin{aligned}
&\delta {\phi}_{laser}=\delta {\phi}_{laser}^{upper}-\delta {\phi}_{laser}^{lower}=\sum_{i}\Delta{\phi}_{i}({\bf x}_{u}(t_{i}),t_{i})-\sum_{i}\Delta{\phi}_{i}({\bf x}_{l}(t_{i}),t_{i})
\end{aligned}
\end{equation}
where the sum is over all of the laser-atom interactions of the lower and upper ensemble. Since the laser pulse travels vertically and the velocity-dependent force is acting in the ${y}$ direction only, the laser phase shift does not depend on the angular velocity of the cylinders. Instead, it manifests itself as a non-zero offset between the two wave packets, and hence may be hard to distinguish from potential systematic errors.\\\indent
It is worth noting that this is only an estimate, and in reality the laser phase shift varies slightly due to presence of shot noise. Shot noise arises as a result of fluctuations in the number of atoms in the ensemble and tends to be the limiting factor to sensitivity. For a typical atom interferometer, the number of atoms in the ensemble is ${n\sim 10^{8}}$, which results in phase sensitivity of ${\sim 1/\sqrt{n}=10^{-4}\mathrm{rad}}$.\\\indent
The propagation phase arises during the "undisturbed" evolution of the ensembles and can be directly computed from the action of a single atom \cite{Book_7}:
\begin{equation}
\delta {\phi}_{propagation}=\frac{\delta S}{\hslash}=\frac{mc^{2}}{\hslash}\left(\int_{0}^{t_{2}}d\tau_{u}-\int_{0}^{t_{2}}d\tau_{l}\right)
\end{equation}
where ${\tau_{u}}$ and ${\tau_{l}}$ are the proper times of the upper and lower ensembles respectively. Using the conservation of energy (20) we have:
\begin{equation}
{\partial}_{t}\tau =c({\partial}_{\tau}x^{0})^{-1}=-\frac{c^{2}(g_{00}^{bg}+h_{00}^{tot})}{\frac{E}{m}+2ch_{02}^{tot}{\partial}_{\tau}x^{2}+2ch_{03}^{tot}{\partial}_{\tau}x^{3}}
\end{equation}
Since ${1>>h_{00}^{tot}>>h_{02}^{tot},h_{03}^{tot}}$, the above expression can be Taylor expanded around ${h_{02}^{tot}=h_{03}^{tot}=0}$. Keeping in mind that ${g_{00}^{bg}=-1-2\frac{{\Phi}^{bg}}{c^{2}}}$ and that in the non-relativistic limit ${E\approx mc^{2}, {\partial}_{\tau}x^{2}\approx {\partial}_{t}x^{2},{\partial}_{\tau}x^{3}\approx {\partial}_{t}x^{3}}$, equation (30) can be to the leading order approximated by:
\begin{equation}
{\partial}_{t}\tau \approx 1+2\frac{{\Phi}^{bg}}{c^{2}}-h_{00}^{tot}-\frac{2}{c}h_{02}^{tot}{\partial}_{\tau}x^{2}-\frac{2}{c}h_{03}^{tot}{\partial}_{\tau}x^{3}
\end{equation}
Substituting (31) into (29), an analytic expression for ${\delta {\phi}_{propagation}}$ is obtained:
\begin{equation}
\begin{aligned}
&\delta {\phi}_{propagation}=\frac{mc^{2}}{\hslash}\int_{0}^{t_{2}}\left(\frac{d\tau_{u}}{dt}-\frac{d\tau_{l}}{dt}\right)dt\\
&=\frac{2m}{\hslash}\int_{0}^{t_{2}}\left({\Phi}^{bg}({\bf x}_{u}(t)+\delta {\bf x}_{u}(t))-{\Phi}^{bg}({\bf x}_{l}(t)+\delta {\bf x}_{l}(t))\right)dt\\
&-\frac{mc^{2}}{\hslash}\int_{0}^{t_{2}}\left(h_{00}^{tot}({\bf x}_{u}(t)+\delta {\bf x}_{u}(t))-h_{00}^{tot}({\bf x}_{l}(t)+\delta {\bf x}_{l}(t))\right)dt\\
&-\frac{2mc}{\hslash}\int_{0}^{t_{2}}\left(h_{02}^{tot}({\bf x}_{u}(t)+\delta {\bf x}_{u}(t)){\partial}_{t}y_{u}(t)-h_{02}^{tot}({\bf x}_{l}(t)+\delta {\bf x}_{l}(t)){\partial}_{t}y_{l}(t)\right)dt\\
&-\frac{2mc}{\hslash}\int_{0}^{t_{2}}\left(h_{03}^{tot}({\bf x}_{u}(t)+\delta {\bf x}_{u}(t)){\partial}_{t}z_{u}(t)-h_{03}^{tot}({\bf x}_{l}(t)+\delta {\bf x}_{l}(t)){\partial}_{t}z_{l}(t)\right)dt
\\
\end{aligned}
\end{equation}
In the above equation, ${m}$ denotes the mass of a single atom, not to be confused with the total mass of the ensemble.\\\indent
The first two terms on the right hand side of (32) are the integrals over ${h_{00}^{tot}}$ and ${{\Phi}^{bg}}$, hence they depend on ${\omega}$ only implicitly through ${\delta {\bf x}_{u}(t)}$ and ${\delta {\bf x}_{l}(t)}$. Because the perturbations ${\delta {\bf x}_{u}(t),\delta {\bf x}_{l}(t)}$ are small, ${h_{00}^{tot}}$ and ${{\Phi}^{bg}}$ may be Taylor expanded to the first order around the unperturbed trajectories ${{\bf x}_{u}(t), {\bf x}_{l}(t)}$:
\begin{equation}
\begin{aligned}
&{\Phi}^{bg}({\bf x}_{u}(t)+\delta {\bf x}_{u}(t))\approx {\Phi}^{bg}({\bf x}_{u}(t))+\nabla{\Phi}^{bg}{\lvert}_{{\bf x}_{u}(t)}\cdot \delta {\bf x}_{u}(t)\\
&{\Phi}^{bg}({\bf x}_{l}(t)+\delta {\bf x}_{l}(t))\approx {\Phi}^{bg}({\bf x}_{l}(t))+\nabla{\Phi}^{bg}{\lvert}_{{\bf x}_{l}(t)}\cdot \delta {\bf x}_{l}(t)\\
&h_{00}^{tot}({\bf x}_{u}(t)+\delta {\bf x}_{u}(t))\approx h_{00}^{tot}({\bf x}_{u}(t))+\nabla h_{00}^{tot}{\lvert}_{{\bf x}_{u}(t)}\cdot \delta {\bf x}_{u}(t)\\
&h_{00}^{tot}({\bf x}_{l}(t)+\delta {\bf x}_{l}(t))\approx h_{00}^{tot}({\bf x}_{l}(t))+\nabla h_{00}^{tot}{\lvert}_{{\bf x}_{l}(t)}\cdot \delta {\bf x}_{l}(t)
\end{aligned}
\end{equation}
However, since the velocity-dependent force acts in the ${y}$ direction, ${\delta {\bf x}_{u}(t)=(0,\delta y_{u},0),\delta {\bf x}_{l}(t)=(0,\delta y_{l},0)}$ have a single non-vanishing component. Therefore, the scalar products reduce to only one component containing derivative with respect to ${y}$. As mentioned before, ${{\Phi}^{bg}}$ and ${h_{00}^{tot}}$ are symmetric around ${y=0}$, hence the derivative vanishes and the contributions of ${{\Phi}^{bg},h_{00}^{tot}}$ to the propagation phase are (to first order in ${\delta {\bf x}_{u}(t), \delta {\bf x}_{l}(t)}$) ${\omega}$ independent.\\\indent
The other contributions to the propagation phase shift come from integrating over the non-diagonal components of the metric ${h_{02}^{tot}}$ and ${h_{03}^{tot}}$. The ${y}$ component of velocity originates solely from the velocity-dependent force ${m{\bf v}\times {\bf B}}$ and hence the first term corresponding to ${h_{02}^{tot}}$ is estimated to be of second order in the non-diagonal components of the metric. As a result, the correction is very small and can be disregarded. The last contribution comes from ${h_{03}^{tot}}$: by expanding to first order in ${\delta {\bf x}_{u}(t),\delta {\bf x}_{l}(t)}$:
\begin{equation}
\begin{aligned}
&\int_{0}^{t_{2}}\left(h_{03}^{tot}({\bf x}_{u}(t)+\delta {\bf x}_{u}(t)){\partial}_{t}z_{u}(t)-h_{03}^{tot}({\bf x}_{l}(t)+\delta {\bf x}_{l}(t)){\partial}_{t}z_{l}(t)\right)dt\\
&\approx \int_{0}^{t_{2}}\left(h_{03}^{tot}({\bf x}_{u}(t)){\partial}_{t}z_{u}(t)-h_{03}^{tot}({\bf x}_{l}(t)){\partial}_{t}z_{l}(t)\right)dt\\
&+\int_{0}^{t_{2}}\left(\nabla h_{03}^{tot}{\lvert}_{{\bf x}_{u}(t)}\cdot \delta {\bf x}_{u}(t){\partial}_{t}z_{u}(t)-\nabla h_{03}^{tot}{\lvert}_{{\bf x}_{l}(t)}\cdot \delta {\bf x}_{l}(t){\partial}_{t}z_{l}(t)\right)dt\\
\end{aligned}
\end{equation}
the second term to the right hand side of the equation (34) is again of second order in non-diagonal components of the metric tensor, and hence may be disregarded. Finally, the last expression:
\begin{equation}
\begin{aligned}
&\int_{0}^{t_{2}}\left(h_{03}^{tot}({\bf x}_{u}(t)){\partial}_{t}z_{u}(t)-h_{03}^{tot}({\bf x}_{l}(t)){\partial}_{t}z_{l}(t)\right)dt\\
&=\int_{0}^{z_{u}(t_{2})}h_{03}^{tot}({\bf x}_{u}(z_{u}))dz_{u}-\int_{0}^{z_{l}(t_{2})}h_{03}^{tot}({\bf x}_{l}(z_{l}))dz_{l}
\end{aligned}
\end{equation}
can be written as a path integral over the classical trajectories of the ensembles. However, since the classical paths ${{\bf x}_{u}(t)=(0,0,z_{u}(t)), \ {\bf x}_{l}(t)=(0,0,z_{l}(t))}$ by design reunite at ${t=t_2}$, the above integral vanishes and the measurable part of the propagation phase is ${\omega}$ independent. As a result, both ${\delta {\phi}_{laser}}$ and ${\delta {\phi}_{propagation}}$ are not of significant importance to our considerations.\\\indent
The only contribution which is dependent on the angular velocity of the discs ${\omega}$ is the separation phase shift. The separation phase depends on the trajectories of the ensembles and hence changes with the angular velocity of the cylinders ${\omega}$. It is given by the scalar product of two vectors:
\begin{equation}
\delta {\phi}_{separation}=\frac{1}{\hslash}{\bf p}\cdot \Delta {\bf x}
\end{equation}
where ${{\bf p}=\frac{m}{2}({\partial}_{t}{\bf x}_{l}{\lvert}_{t_{2}}+{\partial}_{t}{\bf x}_{u}{\lvert}_{t_{2}})}$ is the average momentum of the atoms, while ${\Delta {\bf x}=({\bf x}_{u}+{\delta \bf x}_{u}){\lvert}_{t_{2}}-({\bf x}_{l}+{\delta \bf x}_{l}){\lvert}_{t_{2}}}$ is the separation between the two ensembles arriving at the detector. By design, if ${\omega =0}$ the two ensembles meet at ${t=t_{2}}$ and the separation phase ${\delta {\phi}_{separation}=\frac{1}{\hslash}{\bf p}\cdot ({\bf x}_{u}-{\bf x}_{l}){\lvert}_{t_{2}}}$ vanishes. On the other hand, for rotating cylinders (${\omega \neq 0}$) there is additional force ${m{\bf v}\times {\bf B}}$ acting on the ensembles and they no longer reunite. Since atoms move approximately vertically in presence of ${{\bf B}}$ field directed along ${x}$ axis, the resulting force points in the ${y}$ direction and the separation vector ${\Delta {\bf x}=({\delta \bf x}_{u}-{\delta \bf x}_{l}){\lvert}_{t_{2}}=(0,{\Delta x}_{y},0)}$ has only one non-vanishing component. Therefore, if the assemblies have a non-zero component ${p_{y}}$ of the momentum in the ${y}$ direction when arriving at the detector, the change of ${\omega}$ will cause a change in ${{\Delta x}_{y}}$, hence leading to a potentially measurable correction to ${\delta {\phi}_{separation}}$. The separation phase shift can be expressed in terms of the classical trajectories of the ensembles by rewriting the separation vector ${\Delta {\bf x}}$ as an integral:
\begin{equation}
\begin{aligned}
&\delta {\phi}_{separation}=\frac{1}{\hslash}{\bf p}\cdot \Delta {\bf x}=\frac{1}{\hslash}{\bf p}\cdot (\delta {\bf x}_{u}(t_{2})-\delta {\bf x}_{l}(t_{2}))=\frac{p_{y}}{\hslash}(\delta y_{u}(t_{2})-\delta y_{l}(t_{2}))\\
&=\frac{p_{y}}{\hslash}\int_{0}^{t_{2}}\frac{F_{u}-F_{l}}{m}dt^{2}=\frac{p_{y}}{\hslash}\int_{0}^{t_{2}}\left(B(z_{u}(t)){\partial}_{t}z_{u}(t)-B(z_{l}(t)){\partial}_{t}z_{l}(t)\right)dt^{2}
\end{aligned}
\end{equation}
where equation (22) is used to express the force in terms of ${\bf B}$, which can be calculated from the non-diagonal components of the metric. The separation phase shift is of particular interest, as it is the only leading order ${\omega}$ dependent correction to the classical phase shift.\\\indent
The separation phase shift exhibits a relatively simple dependence on parameters of the experiment. The ${\omega}$ term appearing in ${\delta {\phi}_{separation}}$ is especially important, as the angular velocity of the cylinders can be easily adjusted between different measurements. Since ${{B}_{i}=c{\epsilon}_{ijk}{\partial}_{j}h_{0k}}$ and the non-diagonal components of the metric are linear in ${\omega}$, the rotational velocity of the cylinders may be taken outside of the integral (37). Therefore, the phase shift increases linearly with ${\omega}$.\\\indent
It is instructive to evaluate the ${\omega}$-dependent term ${\delta {\phi}_{separation}}$ for a modern atomic interferometer. Provided the cylinders are made out of steel (steel is chosen for its low cost, heavy weight and high tensile strength that allows it to withstand significant centrifugal forces), it is estimated that the optimal radius and thickness are ${\sim 3\mathrm{m}}$ and ${\sim 0.4\mathrm{m}}$ respectively. Large radius allows for a grater linear velocity near the vacuum chamber of the interferometer (hence, magnifying the frame-dragging effect), while greater thickness increases the total mass and prevents warping of the cylinders. The angular velocity of the discs is limited to ${\omega \sim 2\cdot 10^{2}\mathrm{rad/s}}$, as otherwise the cylinders could be torn apart by the corresponding centrifugal forces. For a regular interferometer of length ${L\sim 10\mathrm{m}}$, the typical interrogation time and changes in velocity ${\Delta v}$ due to the Bragg and Raman scatterings are ${t_{2}=2t_{1}\sim 1.4\mathrm{s}}$ and ${\Delta v\sim 14\mathrm{m/s}}$ respectively. The position of the centres of the cylinders is chosen to be ${y_{c}=3.15\mathrm{m}, z_{c}=4\mathrm{m}}$, such that the cylinders are close to the interferometer's arm and the phase-shift is maximized. Assuming that the ensembles arrive at the detector with a horizontal velocity ${v_{y}\sim 10\mathrm{m/s}}$, the ${\omega}$ dependent part of the phase shift is evaluated numerically to be ${\delta {\phi}_{separation}=-9.3\cdot 10^{-10}\mathrm{rad}}$. The separation phase as a function of ${\omega}$ is shown in Fig.~\ref{fig:fig_5}.
\section{Further research and improvements}
Currently, the most sensitive ${10\mathrm{m}}$ interferometers are capable of measuring phase shifts as low as ${10^{-4}\mathrm{rad}}$. This, however, is still ${5}$ orders of magnitude away from the calculated phase shift of ${\delta {\phi}_{separation}\sim 10^{-9}\mathrm{rad}}$ resulting from the frame-dragging effect of the rotating cylinders. Although the obtained result is not yet within the measuring capabilities of existing aperture, there are numerous improvements possible that have the potential to close the arising gap. The key enhancements are summarized below:
\begin{itemize}
\item Increasing the horizontal momentum ${p_{y}}$ of the the atoms that enter the detector.
  \item Increasing either the length of the interferometer's arm, the interrogation time or the vertical velocity of the ensembles.
  \item Further developing the detection accuracy.
  \item Optimalization of the cylinder design.
  \item Use of more than two cylinders.
  \item Performing multiple measurements.
\end{itemize}
Perhaps one of the easier ways to boost the ${\omega}$ dependent part of the phase shift is to increase the ${y}$ component of the momentum of the ensembles at the end of the interferometer sequence. This can be done by fine-tuning the laser pulse that redirects the beam horizontally towards the detector. Since the separation phase shift increases linearly with the average horizontal momentum ${p_{y}}$ of the two ensembles, increasing the velocity from ${v_{y}\sim 10\mathrm{m/s}}$ to ${v_{y}\sim 100\mathrm{m/s}}$ would result in a tenfold increase in ${\delta {\phi}_{separation}}$.\\\indent
Improving the length of the interferometer's arm, the interrogation time and the vertical velocity of the atoms is a viable option to increase ${\delta {\phi}_{separation}}$. Next-generation atom interferometers are currently being developed, with interrogation time of ${t_{2}\sim 1.4\mathrm{s}}$ and lengths between ${100\mathrm{m}-2000\mathrm{m}}$. The use of such interferometers could mean a two order of magnitude increase in ${\delta {\phi}_{separation}}$.\\\indent
\begin{figure*}
\center
\includegraphics[width=\textwidth]{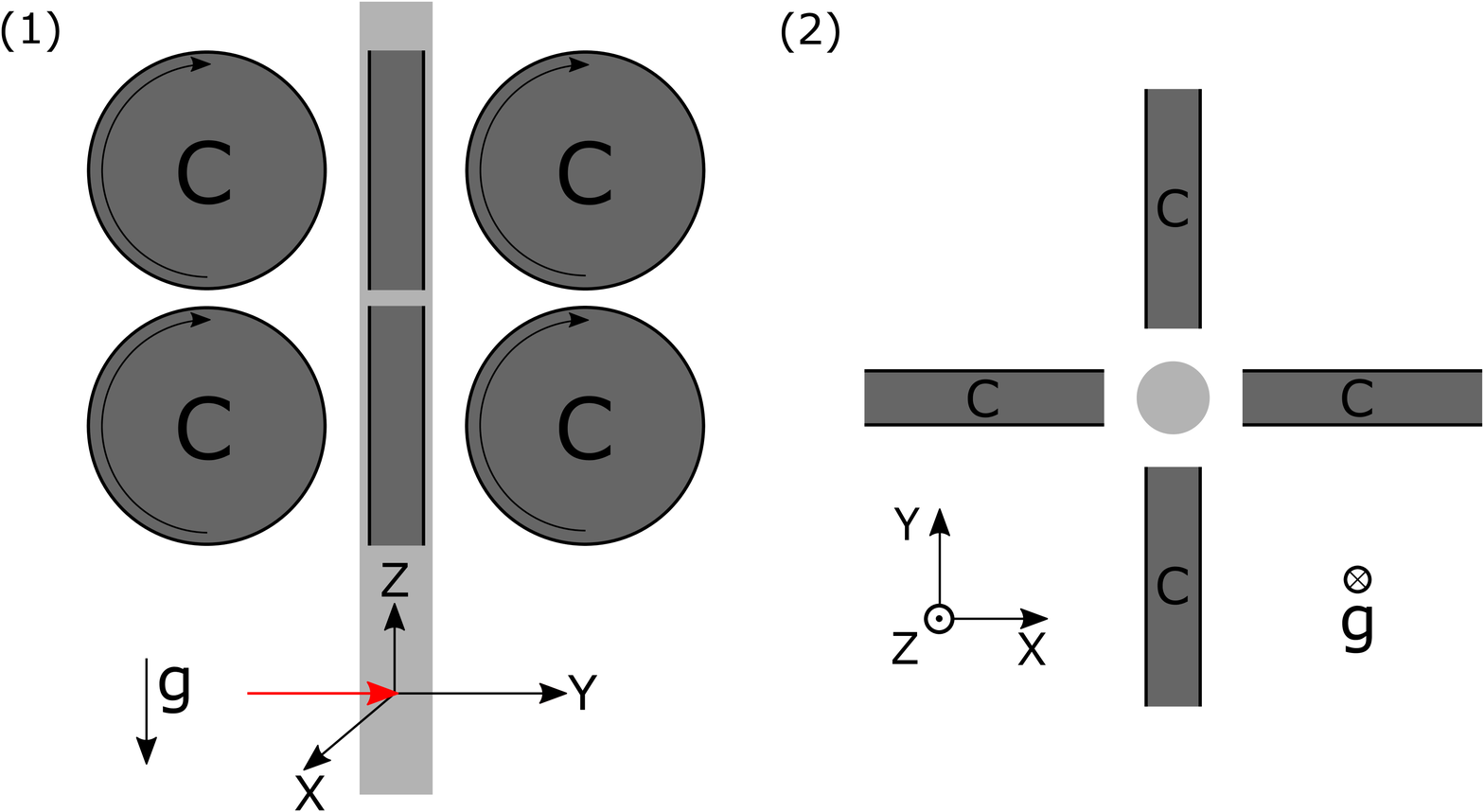}
\caption{\label{fig:fig_6} The disposition of additional cylinders in the improved experimental setup. Cylinders may be placed next to each other both vertically (1) and symmetrically around the interferometer's vacuum chamber when seen from above (2). This does not only magnifies the frame-dragging effect, but also preserves the symmetry of the experimental setup assumed during theoretical calculations. The views (1) and (2) are from the side and from above respectively.}
\end{figure*}
The current goal set for the accuracy of next-generation atom interferometers is around ${10^{-5}\mathrm{rad}}$, with the AION-km reaching an accuracy of up to ${3\cdot 10^{-6}\mathrm{rad}}$. If the predicted accuracy is indeed reached, the current gap will be significantly reduced. For the AION-km project, taking ${t_{2}\sim 1\mathrm{s}}$ and ${{\partial}_{t}z\sim 10^{3}\mathrm{m/s}}$ we obtain ${\delta {\phi}_{separation}\sim 10^{-7}\mathrm{rad}}$. The idea of using new-generation interferometers is very promising, when comparing ${\delta {\phi}_{separation}\sim 10^{-7}\mathrm{rad}}$ with the predicted sensitivity of ${3\cdot 10^{-6}\mathrm{rad}}$.\\\indent
In the article, two symmetrically placed cylinders were used as a model for the frame-dragging effect generator. Although it provided an instructive example and greatly simplified the calculations, the cylinder is not the most optimal shape for this purpose. The goal is to maximize ${h_{02}^{tot}, h_{03}^{tot}}$, while for practical reasons keeping the mass and volume of the cylinder bounded. Since the non-diagonal parts of the metric depend linearly on ${\omega}$, the stresses on the rotating body have to be taken into account. Furthermore, steel may not be the best choice of material. Possibly, mass should be displaced toward the outer edge, as ${h_{02}^{tot}, h_{03}^{tot}}$ increase rapidly with the radius of a rotating body.\\\indent
Instead of using only two cylinders, several can be employed in a symmetrical arrangement around the interferometer's vacuum chamber to magnify the gravitational effect. In the case of a longer interferometer, it would be both feasible and necessary, as to maintain the frame-dragging metric along the chamber. Proposed arrangement of added cylinders is presented in Fig.~\ref{fig:fig_6}.\\\indent
Finally, numerous measurements can be taken as to statistically improve accuracy of the experiment. As opposed to detection of gravitational waves, the gravitational effects due to rotating bodies are consistent, and can be maintained over an extended period of time.
\section{Conclusion}
The main goal of the above analysis was to show that atom interferometers, previously considered primarily for detection of dark matter and gravitational waves, are a viable mean of probing the small-scale regime of general relativity. As a proof of concept, we proposed a simple experiment constituting an atom interferometer and set of massive rapidly-rotating cylinders. The gravitational field around a spinning cylinder is considered, and the leading order formula for the phase shift is given. The calculations performed can be easily generalized to any axisymmetric rotating body. The phase shift is then estimated in the case of a modern interferometer.\\\indent
Numerical calculations indicate a difference of the order of ${10^{-5}\mathrm{rad}}$ between the generated phase shift and sensitivity of modern interferometers. Although it may seem substantial, the field of accurate atom interferometry is still in its development, with significant progress foreseen in the coming decade. It is therefore expected that for the new generation of large-scale atom interferometers, the discrepancy would be within a range of order ${10^{-1}\mathrm{rad}}$ to ${10^{-2}\mathrm{rad}}$. In addition,  various improvements to the experimental setup were proposed, possibly closing the gap and resulting in the first measurements of a purely general-relativistic effects over small distances.

\bmhead{Acknowledgments}
I would like to show my deep appreciation to Prof. Pat Roche and Prof. Christopher Foot for valuable remarks and suggestions. I also wish to acknowledge the help provided by Reemon Spector on earlier versions of the manuscript.

\begin{appendices}

\section{Einstein-Maxwell formalism}\label{secA1}

The Einstein-Maxwell linearized gravity is a first order approximation of geodesic equations that, unlike the Newtonian formalism, takes into account the velocity-dependent forces. The approximation remains valid when examining the trajectories of slow-moving test particles affected by stationary (time independent) non-relativistic sources. In this section, brief explanation of the method is given, while further information can be found e.g. \cite{Book_9}.\\\indent
Let the linearized metric in Cartesian coordinates be written as:
\begin{equation}
g_{\mu \nu}={\eta}_{\mu \nu}+h_{\mu \nu},
\end{equation}
where ${{\eta}_{\mu \nu}=diag[-1,1,1,1]}$ is the flat background metric, and ${{h}_{\mu \nu}}$ is the perturbation term. For a non-relativistic particle (particle for which ${v<<c}$), the position four-vector:
\begin{equation}
x^{\mu}=(ct,{\bf x})
\end{equation}
can be differentiated with respect to the proper time ${\tau}$ to give the four-velocity:
\begin{equation}
u^{\mu}={\gamma}_{\bf v}(c,{\bf v})\approx (c,{\bf v}).
\end{equation}
Because ${{\gamma}_{\bf v}\approx 1}$, the proper time in the geodesic equations can be replaced with the absolute time ${\tau \approx t}$. The geodesic equations read:
\begin{equation}
{\partial}_{\tau}^{2}x^{\mu}=-{\Gamma}_{\alpha \beta}^{\mu}{\partial}_{\tau}x^{\alpha}{\partial}_{\tau}x^{\beta}.
\end{equation}
Keeping only the leading terms in ${v/c}$, the spatial part of the above equations reduces to:
\begin{equation}
{\partial}_{t}^{2}x^{i}\approx -c^{2}\left({\Gamma}_{00}^{i}+2{\Gamma}_{0j}^{i}\frac{v^{j}}{c}+{\Gamma}_{jk}^{i}\frac{v^{j}v^{k}}{c^{2}}\right)\approx -c^{2}{\Gamma}_{00}^{i}-2c{\Gamma}_{0j}^{i}v^{j}
\end{equation}
where the Latin indices run over spatial components, i.e. ${i,j,k=1,2,3}$.\\\indent
If the terms that are linear in velocity ${{\Gamma}_{0j}^{i}v^{j}}$ are disregarded, the regular Newtonian equation is retrieved. However, by keeping the second terms on the right hand side of (A5), one can rewrite the dynamical equations in analogy to the Lorentz equation in electrodynamics. The Christoffel symbols expanded to the leading order in ${{h}_{\mu \nu}}$ are:
\begin{equation}
{\Gamma}_{00}^{i}=-\frac{1}{2}{\partial}^{i}h_{00}, \ {\Gamma}_{0j}^{i}=\frac{1}{2}\left({\partial}_{j}h_{0}^{i}-{\partial}^{i}h_{0j}\right).
\end{equation}
The above formulas can be now substituted into (A5), simplifying to:
\begin{equation}
{\partial}_{t}^{2}x^{i}=\frac{1}{2}c^{2}{\partial}^{i}h_{00}+c\left({\partial}^{i}h_{0j}-{\partial}_{j}h_{0}^{i}\right)v^{j}.
\end{equation}
In order to recast the perturbed geodesic equations into a more familiar form, it is convenient to work in terms of auxiliary 3D fields. If the two gravitational vector fields are defined in terms of the metric as:
\begin{equation}
E_{i}=\frac{c^{2}}{2}{\partial}_{i}{h}_{00}, \ {B}_{i}=c{\epsilon}_{ijk}{\partial}_{j}h_{0k},
\end{equation}
then the equations (A7) can be rewritten in the form:
\begin{equation}
{\partial}_{t}^{2}{\bf x}={\bf E}+{\bf v}\times {\bf B},
\end{equation}
which clearly resembles the Lorentz force for electromagnetism and, as opposed to Newton's equation, includes the velocity-dependent term ${{\bf v}\times {\bf B}}$.\\\indent
The result (A9) is simply a more inclusive version of the Newton's equation. It remains valid, as long as the test particles move non-relativistically in a time-independent gravitational field. In addition, only the linear terms in the perturbed metric were kept during the derivation, hence the weak-field limit is assumed to hold as well.
\end{appendices}

\section*{Declarations}
Data sharing not applicable to this article as no datasets were generated or analysed during the current study. Furthermore, this manuscript has not been published and is not under consideration for publication elsewhere. We have no conflicts of interest to disclose.

\bibliography{sn-bibliography}


\end{document}